\documentclass[12pt]{article}
\usepackage{amssymb,amsmath,graphicx}\pdfoutput=1
\begin{document}
\title{\textbf{Large $N$ flavor $\beta$-functions: a recap}} 
\author{B.~Holdom%
\thanks{bob.holdom@utoronto.ca}\\
\emph{\small Department of Physics, University of Toronto}\\[-1ex]
\emph{\small Toronto ON Canada M5S1A7}}
\date{}
\maketitle
\begin{abstract}
$\beta$-functions for abelian and non-abelian gauge theories are studied in the regime where the large $N$ flavor expansion is applicable. The first nontrivial order in the $1/N$ expansion is known for any value of $N\alpha$, and there are also various indications as to the nature of higher order effects. The singularity structure as a function of $N\alpha$ has implications for the existence of nontrivial fixed points.
\end{abstract}

For a sufficiently large number of flavors a non-abelian gauge theory will loose asymptotic freedom and will in this way resemble an abelian gauge theory. We shall thus focus our attention on the abelian case and then later extend the discussion to the nonabelian case.

It is generally believed that a $U(1)$ gauge theory with $N$ charged fermions has a running coupling that grows monotonically towards the ultraviolet, and thus suffers from a Landau pole. This is the indication from the one-loop $\beta$-function. But there is much more known about the perturbative $\beta$-function and there have been recent calculations that have extended our knowledge to 5-loops. We also have complete knowledge of the first nontrivial order in the $1/N$ expansion for any value of $N\alpha$, and so this makes the large $N$ expansion a useful way to organize the perturbative expansion. The question is whether any of this allows us to glean anything further about the possible existence of nontrivial fixed points. Although any perturbative approach will introduce a renormalization scheme dependence, it can still be hoped that the existence or nonexistence of a fixed point will leave some mark on the perturbative results.

According to a lattice result \cite{a4} there is no nontrivial fixed point in a $U(1)$ gauge theory for $N=4$. We shall be concerned with larger $N$ where the large $N$ expansion suggests other possibilities. This provides some motivation to study larger $N$ values on the lattice as well, for sufficiently large values of $N\alpha$. Extending the current lattice result to $N=8$, 12, 16,... would appear to be relatively straightforward.

The $U(1)$ $\beta$-function is defined as
\begin{equation}
\beta(\alpha)=\frac{\partial \ln \alpha}{\partial \ln \mu}.
\label{e10}\end{equation}
The one loop result is $\beta(\alpha)=2A/3$ where $A\equiv N\alpha/\pi$. We may write  an expansion in $1/N$ as follows,
\begin{equation}
\frac{3}{2}\frac{\beta(\alpha)}{A}=1+\sum_{i=1}^{\infty}\frac{F_i(A)}{N^i}.
\label{e2}\end{equation}
The ``1'' corresponds to the one loop result and we shall refer to it as the zeroth order term in the $1/N$ expansion. Each $F_i(A)$ represents a class of diagrams having the same dependence on $N$ when $A$ is held fixed, and such diagrams exist to all orders in $A$. If the functions $|F_i(A)|$ were bounded then for sufficiently large $N$ one could conclude that the zeroth order term dominates  and that the Landau pole is unavoidable. But singularities in the $F_i(A)$ will keep us from drawing this conclusion.

We collect together what is known about the $F_i(A)$'s in the $\overline{\rm{MS}}$ renormalization scheme. \begin{eqnarray}
F_1(A)&=&\int_0^\frac{A}{3}I_1(x)dx\label{e1}
\\
&&I_1(x)={\frac {  \left( 1+x \right) \left( 2\,x-1 \right)^2  \left( 2\,x-3 \right)^2 \sin \left( \pi \,x \right)^{3} \Gamma 
 \left( x-1 \right)^{2}\Gamma  \left( -2\,x \right) }{ \left( x-2 \right) {\pi }^{3}}}
\label{e9}\\
F_2(A)&=&-{\frac {3}{32}}\,{A}^{2}+ \left( {\frac {95}{288}}-{\frac {13}{12}}\zeta  \left( 3 \right)  \right) {A}^{3}+\left({{
 4961}\over{13824}}+{{11\pi^4}\over{2880}}-{{119\zeta\left(3\right)}\over{144}}\right)A^4+...
\\
F_3(A)&=&-\frac{69}{128}A^3+...
\\
F_4(A)&=& \left( {\frac {4157}{2048}}+\frac{3}{8}\zeta  \left( 3 \right)  \right)A^4+...
\label{e3}\end{eqnarray}
Important for our study is the fact that $F_1(A)$ is known completely \cite{a1}. We have expressed the integrand $I_1(x)$ in a form that makes more clear the location of its zeros and poles. The $A^3$ terms in $F_2(A)$ and $F_3(A)$ were calculated in \cite{a2}, the $A^4$ term in $F_2(A)$ in \cite{a5}, and the $F_4(A)$ term in \cite{a3}. The latter two results are 5-loop calculations.
\begin{figure}[t]
\centering\includegraphics[scale=0.5]{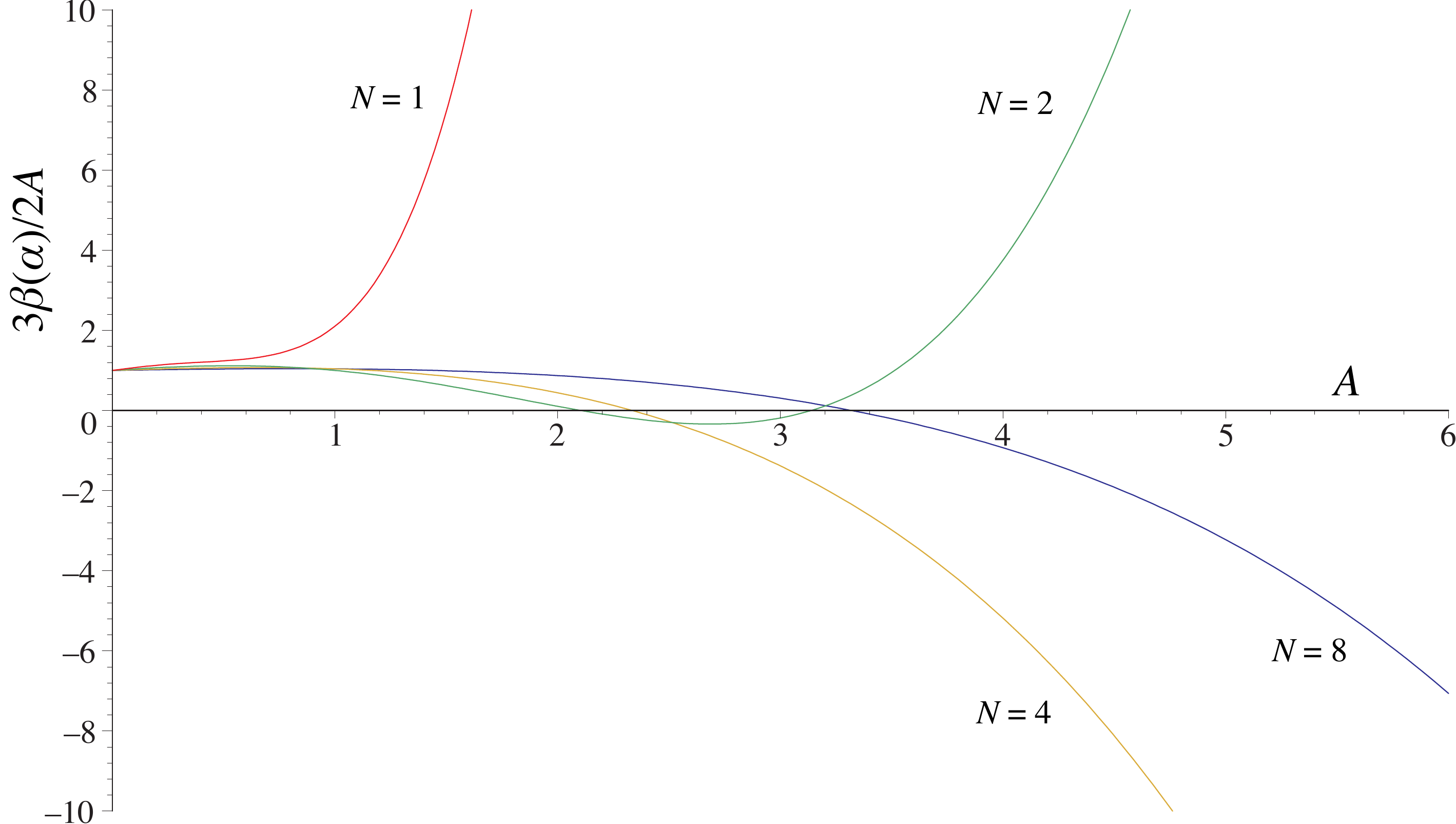}
\caption{The sum of the known contributions to $3\beta(\alpha)/2A$ in (\ref{e1}-\ref{e3}).}\end{figure}

One way to express the results in (\ref{e1}-\ref{e3}) is to plot their sum and ignore what is not known. The result for the $3\beta(\alpha)/2A$ for various $N$ is displayed in Fig.~(1). A zero would indicate a nontrivial fixed point, but the zeros are occurring at values of $A$ that are too high to ignore higher order terms. Thus we cannot deduce much from this plot, except to notice sensitivity of the $\beta$-function to $N$.
\begin{figure}[t]
\centering\includegraphics[scale=0.5]{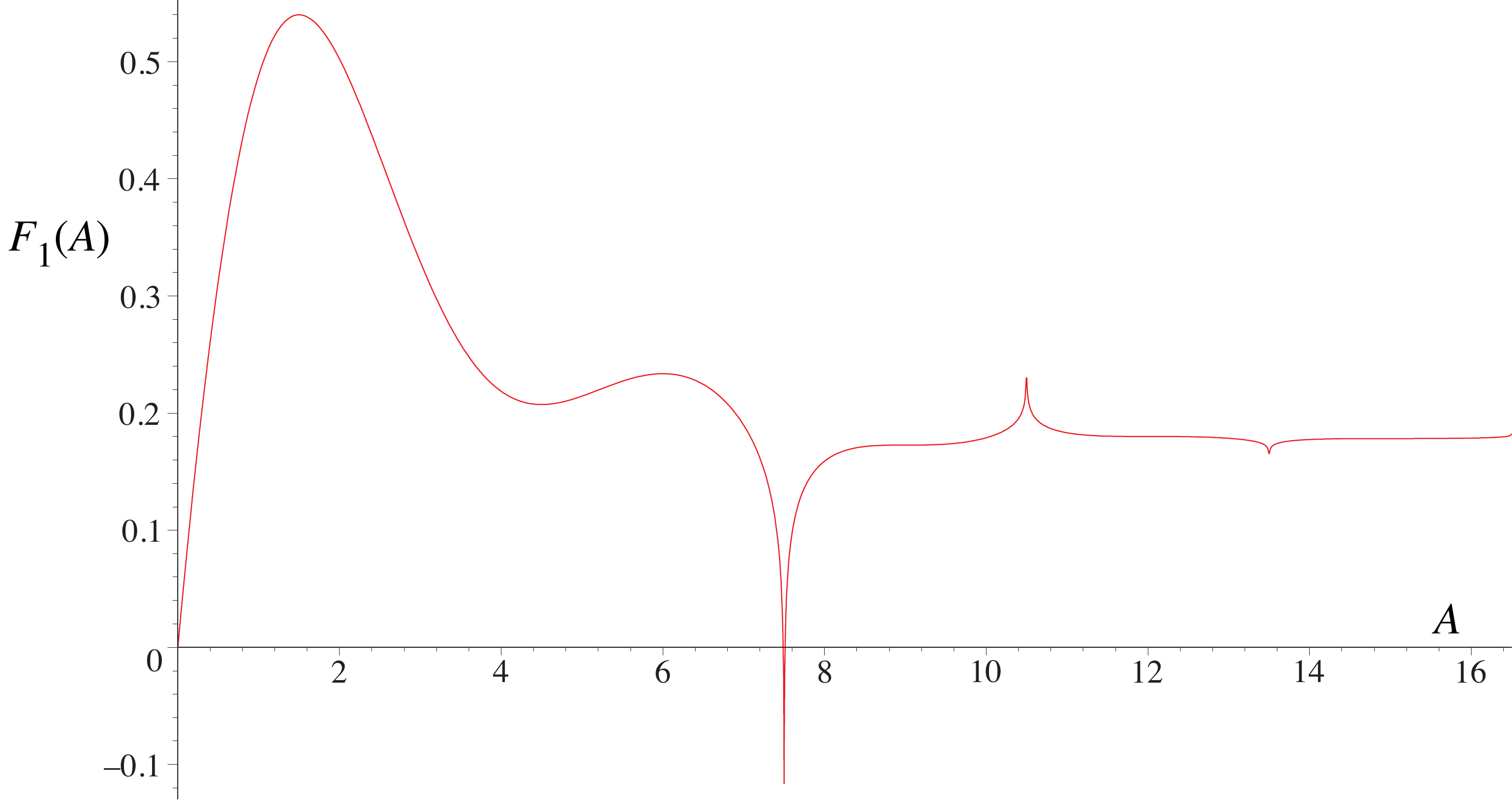}
\caption{$F_1(A)$ as defined in (\ref{e1}).}\end{figure}

The 2-loop contribution to $\beta(\alpha)$ involves one fermion loop and one internal photon and it gives rise to the first term in the expansion of $F_1(A)$, which is $\frac{3}{4}A$. The higher order terms in $F_1(A)$ correspond to the insertion of the appropriate number of fermion loops on the photon line, and these bubble chains have been summed up to produce the result (\ref{e1}) \cite{a1}. An important feature of $F_1(A)$ is that its expansion displays a finite radius of convergence, which is nonvanishing due to the slower than factorial growth in the number of diagrams. We expect this to be true of the other $F_i(A)$'s as well.

Simple poles of alternating sign appear in $I_1(x)$ at $x=\frac{5}{2}+n$ for integer $n\ge0$. The integration can be handled with a Cauchy principal value prescription and the result is shown in Fig.~(2).  Clearly $F_1(A)$ only changes logarithmically as the singular points are approached. And these singularities become weaker for larger $n$. Close to the first singularity at $A=15/2$ we find
\begin{equation}
F_1(A)\approx\frac{7}{15\pi^2}\Re(\log(1-\frac{2}{15}A))+0.3056
.\end{equation}
Nevertheless even this weak singularity can cause the $\beta$-function at this order to vanish at a fixed value of $N$. As Fig.~(2) indicates there will be two nearly coincident zeros of $1+F_1(A)/N$ at
\begin{equation}
A=\frac{15}{2}\pm 0.0117e^{-15\pi^2N/7}
\label{e12}.\end{equation}
The lower (upper) one is an ultraviolet (infrared) fixed point. In either case the running coupling achieves its fixed point value at some finite scale $\mu_*$, and at this scale the running of the coupling abruptly stops.

It is useful to compare the $\beta$-function to the $\gamma_m$-function defined as
\begin{equation}
\gamma_m(\alpha)=-\frac{\partial \ln m}{\partial \ln \mu}=\sum_{i=1}^{\infty}\frac{G_i(A)}{N^i}.
\end{equation}
$G_1(A)$ is also known to all orders in $A$, and in fact it is directly related to $F_1(A)$. From the results of \cite{a1} one can deduce that
\begin{equation}
\frac{dF_1(A)}{dA}=\frac{1}{2A}(1-\frac{2A}{3})(1+\frac{A}{3})G_1(A).
\label{e4}\end{equation}
Thus the singularities in these two functions occur at the same locations. This could be expected since the bubble chain re-summation is intrinsic to both functions. The strength of the singularities are also related; the logarithmic singularities in $F_1(A)$ correspond to simple poles in $G_1(A)$.

It could be argued that the Cauchy principal value prescription used in the evaluation of $F_1(A)$ is not unique. Rather than approaching the pole equally closely from the two sides, another prescription would be to approach the pole unequally from the two sides. This would shift the $\beta$-function on the right of a singularity by an additive constant, as allowed by (\ref{e4}). But this ambiguity does not alter the appearance of the fixed points at first order in $1/N$.

It is important to know how large $N$ has to be for the $1/N$ expansion to be under control. We can require that the known expansion terms of the higher $F_i(A)$'s be sufficiently small for $A$ as large as the radius of convergence of $F_1(A)$, at $A=15/2$. If we rescale $A=15/2\tilde{A}$ and $N=16\tilde{N}$ then the expansion (\ref{e2}) numerically reads
\begin{eqnarray}
1&+&\frac{1}{\tilde{N}}(.3516\tilde{A}-.8057\tilde{A}^2-1.567\tilde{A}^3+5.342\tilde{A}^4+1.60\tilde{A}^5-15.9\tilde{A}^6+...)\nonumber\\
&+&\frac{1}{\tilde{N}^2}(-.0206\tilde{A}^2-1.602\tilde{A}^3-3.244\tilde{A}^4+...)\label{e11}\\
&+&\frac{1}{\tilde{N}^3}(-.0555\tilde{A}^3+...)+\frac{1}{\tilde{N}^4}(.1198\tilde{A}^4+...)\nonumber
\end{eqnarray}
We see that $\tilde{N}\gtrsim1$ or more is needed for the terms of successively higher powers of $1/\tilde{N}$ to be under control. One can in particular compare the leading terms at each order in $1/\tilde{N}$, i.e.~the $\tilde{A}^i$ term at order $1/\tilde{N}^i$. These terms correspond to the one fermion loop diagrams, and they have a special significance in that they are renormalization scheme independent \cite{a2}. 

Thus for sufficiently large $N$ the higher orders in $1/N$ are under control in the usual sense of an asymptotic series. But the presence of singularities in the $F_i(A)$ indicates that the $1/N$ expansion needs to be reconsidered for $A$ close to these singularities. Although the singularities of $F_1(A)$ are logarithmic, we do not expect this to continue for the higher $F_i(A)$. The appearance of poles in $G_1(A)$ reinforces this view. In particular if $F_2(A)$ has a simple pole at $A=15/2$ then this should completely dominate the logarithmic singularity of $F_1(A)$, as the exponentially small spacing in (\ref{e12}) makes clear.

It is interesting that the first three coefficients in the expansion of $F_2(\tilde{A})$ are negative. This is certainly consistent with a pole, since the expansion of the expression
\begin{equation}
\tilde{A}^2\frac{0.771\tilde{A}+0.010}{\tilde{A}-0.494}
\label{e5}\end{equation}
has the same first three terms. But we would argue that this does little to prove the existence of a pole. In the $F_2(A)$ expansion we note that the 2 fermion loop $A^3$ term in $F_2(A)$ receives contributions from graphs with topology different from the 1 fermion loop $A^2$ term. That is, unlike the case of $F_1(A)$, dressing photon lines of the $A^2$ graphs does not give all $A^3$ graphs. In fact the graphs with the new topology (the light-by-light scattering contributions) give the dominant contribution to the $A^3$ term \cite{a2}. In this situation we do not expect a Pade approximant to be very predictive until more orders in the expansion are known.  

An example of the appearance of a pole in an exact $\beta$-function is provided by $SU(N_c)$ SUSY pure gluodynamics where \cite{a7,a8}
\begin{equation}
\beta(a)=\frac{3a^2}{4}\frac{1}{a-2/N_c},\quad a\equiv\frac{\alpha_c}{\pi}
\label{e6}.\end{equation}
A pole of this sign, as in (\ref{e5}), means that the coupling will evolve towards the singularity in the infrared, and it does so from both the weak and strong coupling sides of the pole. The coupling reaches the singularity at some minimum finite renormalization scale $\mu_*$, and the gauge theory ceases to provide a description below this scale. This suggests that the theory develops a mass gap and/or some fundamentally different description is needed for energies below $\mu_*$. The authors of \cite{a9} argue that there is evidence of such a pole in the QCD $\beta$-function from the study of Pade approximants. A pole of opposite sign would instead produce an ultraviolet cutoff on the gauge theory description.

Beyond the $A^3$ term in $F_2(A)$, including the known $A^4$ term, the diagrams \textit{are} obtained by simply inserting fermion loops into photon lines, thus building up bubble chains. This is true for sufficiently high powers of $A$ in the expansion of any $F_i(A)$; one ends up only dressing photons lines with fermion loops in graphs that belong to a basic set of topologies. It is for this reason that all these functions should have a nonvanishing radius of convergence, and the bubble chains should generate singularities in the higher $F_i(A)$ just as they did for $F_1(A)$ and $G_1(A)$. We expect that the set of locations of singularities in the higher $F_i(A)$ will include the locations of singularities in $F_1(A)$.
\begin{figure}[t]
\centering\includegraphics[scale=0.27]{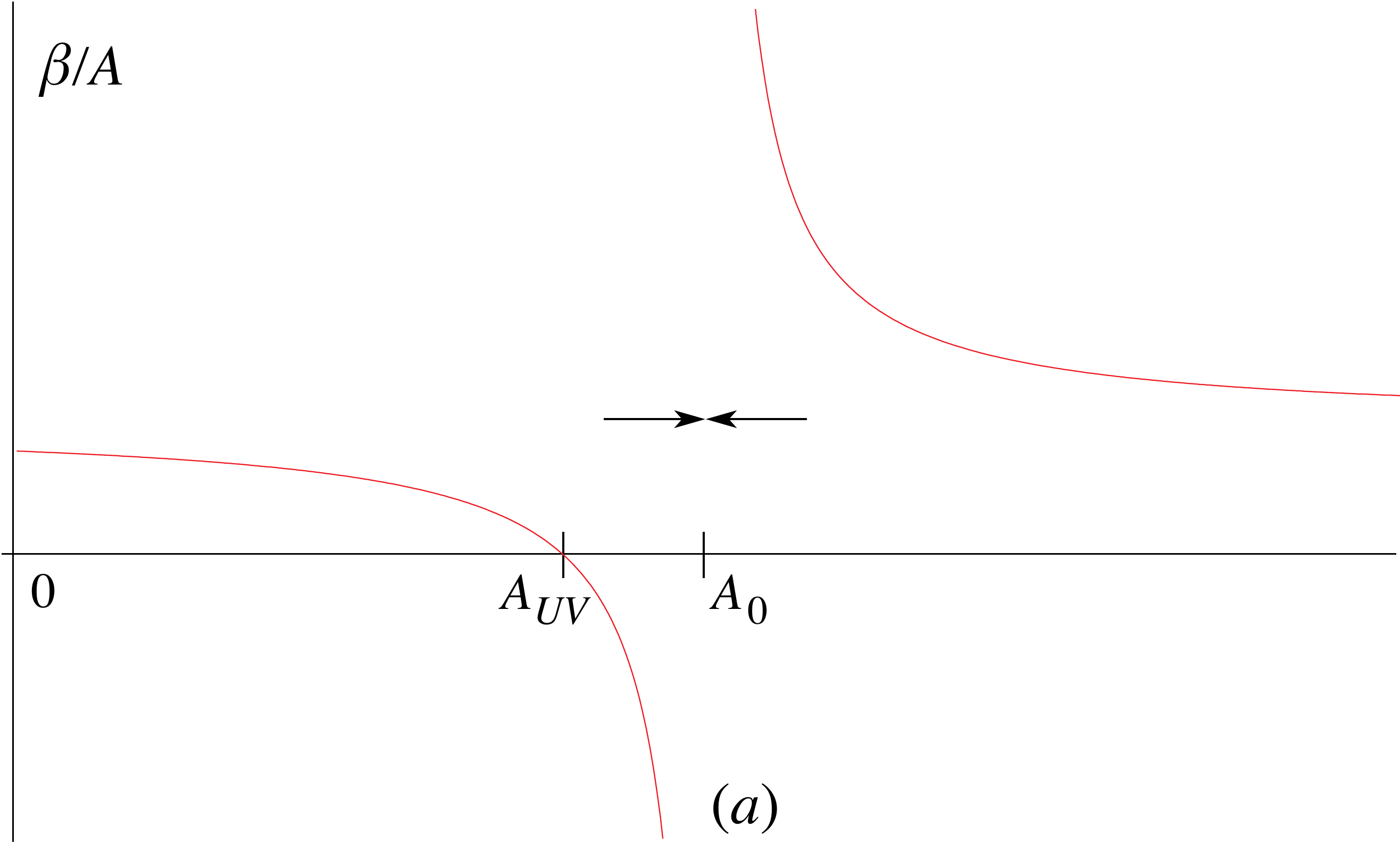}\hspace{7ex}\includegraphics[scale=0.27]{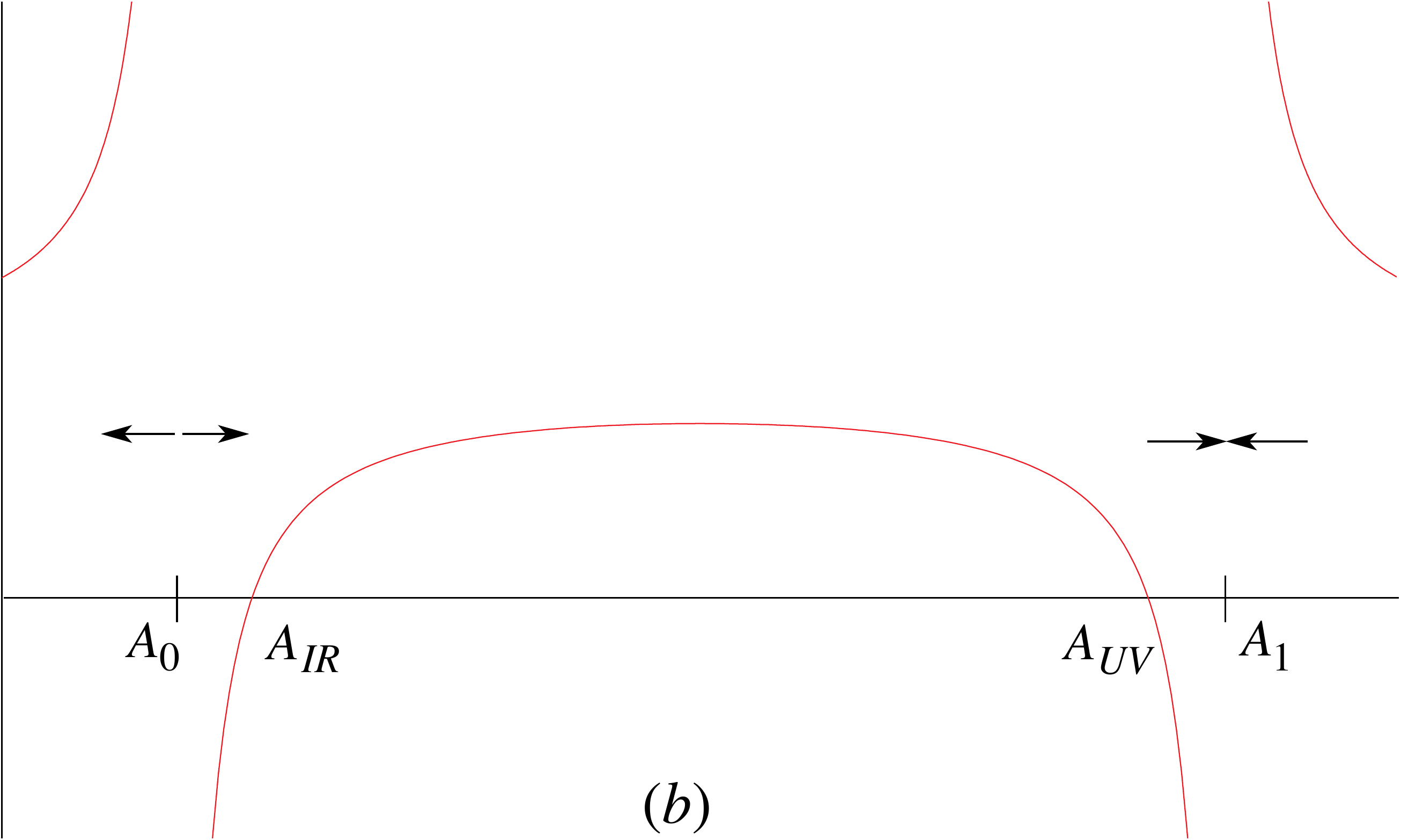}\\
\vspace{2ex}
\centering\includegraphics[scale=0.27]{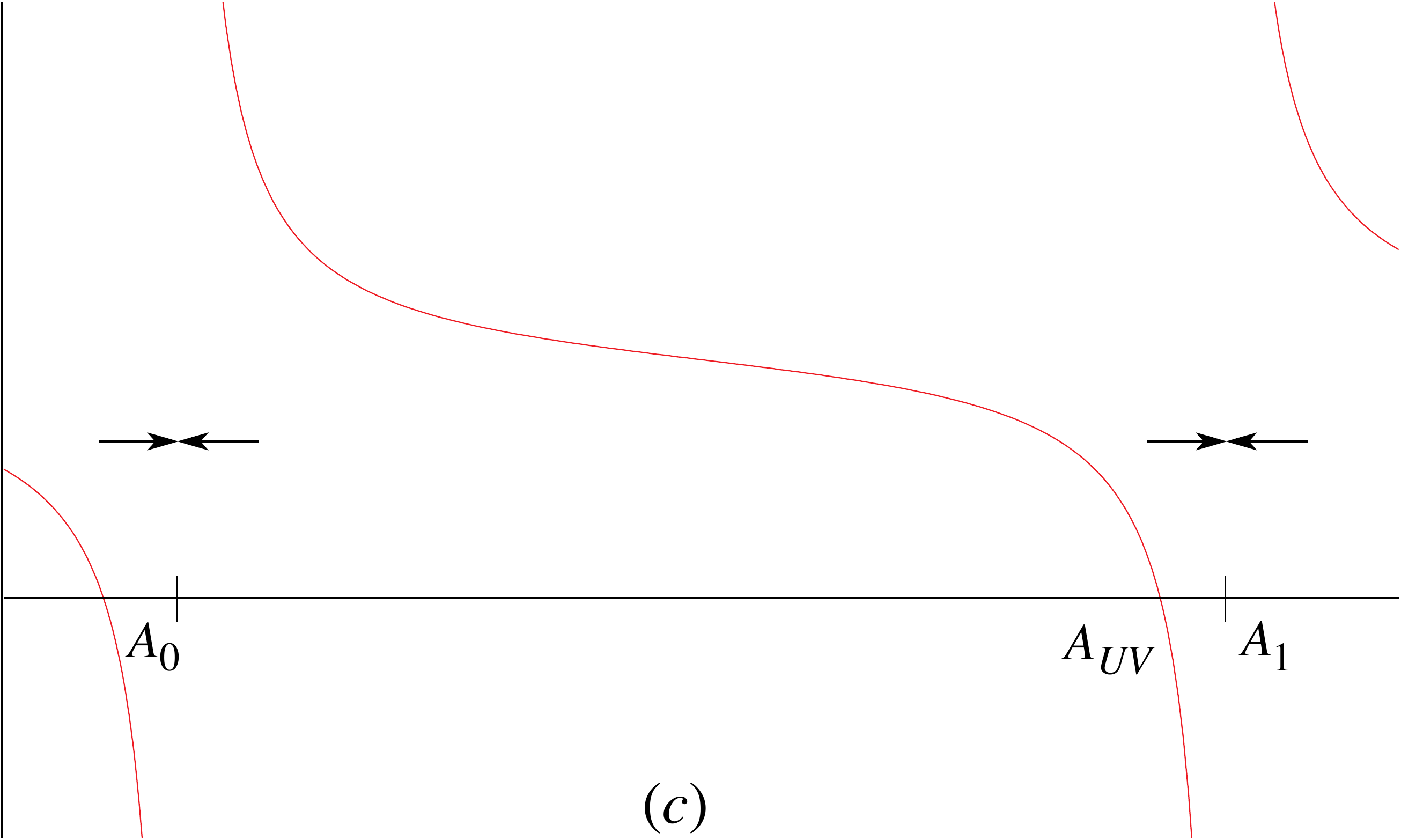}\hspace{7ex}\includegraphics[scale=0.27]{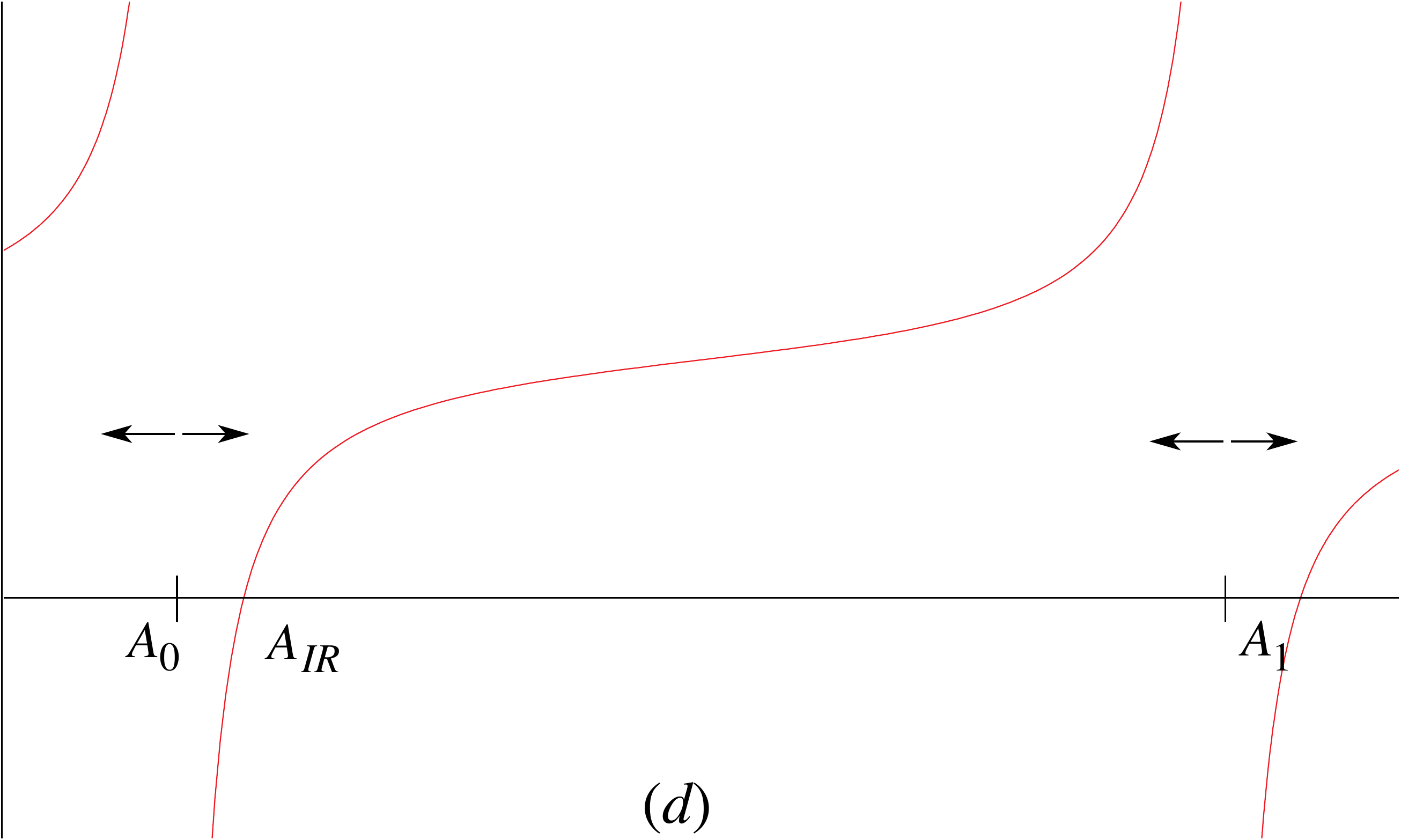}
\caption{Schematic examples of the behavior of $3\beta/2A$ when poles are present. The arrows show the infrared flow close to the poles.}\end{figure}

Let us consider a set of poles in $F_2(A)$ occurring at the same locations at the singularities in $F_1(A)$ (we shall provide more evidence for this shortly). Poles in $F_2(A)$ at $A_n=\frac{15}{2}+3n$ need to be considered along with the zeroth order term in (\ref{e2}). Depending on the sign of these poles, the result is the existence of various nontrivial infrared and/or ultraviolet fixed points. For example an infrared pole at $A_0=15/2$, i.e.~having the sign of $1/(A-A_0)$ as in (\ref{e5}) and (\ref{e6}), implies an ultraviolet fixed point at some $A_{UV}<A_0$ (see Fig.~(3a)). When the coupling is below $A_{UV}$ the theory flows to the known weakly coupled behavior in the infrared, but in the ultraviolet the Landau pole has been eliminated in favor of a fixed point. Possibilities for nontrivial fixed points in the regions between successive poles at $A_i$ and $A_{i+1}$ are shown in Figs.~(3b,c,d).

We now describe some evidence that is relevant not only to the existence of poles in $F_2(A)$, but also to their location and sign. The first input comes from the $\gamma_m$-function, since it turns out that $G_2(A)$ is also known to all orders in $A$. This comes from an impressive calculation in QCD \cite{a13}, from which the corresponding QED result can be extracted. The surprisingly simple result is that the singularities in $G_2(A)$ occur at the same locations as in $G_1(A)$, and each simple pole in $G_1(A)$ has been replaced by a double pole in $G_2(A)$.

Further input comes from examples of theories where the analogs of $G_2(A)$ and $F'_2(A)$ have the same singularity structure; that is the set of poles of these functions have the same location, sign and order of pole. In particular in $N$ flavor $(\phi^2)^2$ theory in $d=4-2\varepsilon$ dimensions the critical exponents $\lambda(\varepsilon)$ and $\omega(\varepsilon)$ have a correspondence to $1-\gamma_m$ and $2\beta'$. In \cite{a12} these critical exponents are encoded in the corresponding critical exponents of the large $N$ $\sigma$-model in $d=2-2\varepsilon$ dimensions, where the functions appearing at order $1/N^i$ are labeled $\lambda_i(\varepsilon)$ and $\omega_i(\varepsilon)$. These functions are the analogs of $-G_i(A)$ and $F'_i(A)$.  From the results in \cite{a12} the functions $-\lambda_2(\varepsilon)$ and $\omega_2(\varepsilon)$ have singularities that match in location, sign and order of pole, as do $-\lambda_1(\varepsilon)$ and $\omega_1(\varepsilon)$.\footnote{$\omega_1(\varepsilon)$ has simple poles, while $\omega_2(\varepsilon)$ has an additional pole and all its poles are fourth order.}

Thus it would not be surprising for QED to display similar behavior, so that $G_2(A)$ and $F'_2(A)$ exhibit a common singularity structure just as $G_1(A)$ and $F'_1(A)$ do. Such behavior could be expected due to the simple relationship between the graphs contributing to $F_i(A)$ and $G_i(A)$, suggesting that the behavior also extends to higher orders. Cutting out an external gauge field vertex from any graph contributing to $F_i(A)$ (vacuum polarization graphs) and replacing the other external vertex with a mass insertion gives a graph contributing to $G_i(A)$ (mass renormalization graphs). This is not true in QCD due to the gluon self-interactions, and so there is no corresponding behavior for QCD even at first order in $1/N$.

Thus supposing that $F'_2(A)$ and $G_2(A)$ have the same singularity structure, the signs of the poles in $F_2(A)$ at $A_n=\frac{15}{2}+3n$ are determined by the known results \cite{a13} for $G_2(A)$. This would imply that the poles in $F_2(A)$ all have the same sign and that they are all ultraviolet poles, i.e.~opposite in sign to (\ref{e5}) and (\ref{e6}). Thus of the various possibilities displayed in Fig.~(3), the available evidence suggests that only Fig.~(3d) is correct for the interval between any two adjacent singularities.

If this is correct then at this order in the $1/N$ expansion there is an infinite number of new theories each flowing towards an associated nontrivial infrared fixed point. There is still the weakly coupled theory that flows to the trivial infrared fixed point. All the theories are cut off in the ultraviolet by the UV poles. We also note that for large $N$ some of the nontrivial infrared fixed points can occur at values of $A$ smaller than the critical value for chiral symmetry breaking, $A_{\rm crit}=N/3$.

To proceed to higher order we must once again reconsider the $1/N$ expansion close to the singularities. As a singularity is approached the $\beta$-function experiences a significant change, and a change of $3\beta/2A$ of order unity would be an indication that the $1/N$ expansion is breaking down for these values of $A$. The higher $F_i(A)$'s must be considered in these regions. For these functions to overcome the suppressions from large $N$ and loop factors, they would need to have progressively higher order poles.

In fact it is plausible that the $F_i(A)$ function has $(i-1)$th order poles, since this behavior is already suggested by the increasing order of the poles in the $G_1(A)$ and $G_2(A)$ functions. A $1/N$ expansion with poles of ever increasing order would indicate the existence of a set of essential singularities in the $\beta$-function. But the implications for fixed points remain similar to our discussion of simple poles. Whenever an essential singularity drives $3\beta/2A$ by an amount of order unity or more in the right direction then a nontrivial fixed point can result.

We may speculate further about the nature of the higher order poles based on an emerging pattern for poles of even or odd order. The adjacent logarithmic (even order) singularities in $F_1(A)$ are alternating in sign while the simple (odd order) poles in $F_2(A)$ are all of the same sign. Let us consider the continuation of this pattern (alternating even order poles and same sign odd order poles) to higher orders, where we treat the overall sign at any order as unknown. It means that on a given interval between adjacent singularities, and at any given order, $3\beta/2A$ diverges with opposite signs on the two ends of the interval. The divergences of the all-orders-summed result (assuming that the summed result diverges) can then also be expected to come with opposite signs on the two ends. This would imply at least one nontrivial ultraviolet or infrared fixed point on each such interval.  Also, the sum of the odd order poles produces a divergence pattern that is the same on every interval, while for the sum of the even order poles the pattern flips in sign on adjacent intervals. Thus the existence of infrared fixed points in the total result, at least on every second interval, is made more likely if the odd poles are always of the ultraviolet variety as in Fig.~(3d). Due to cancellations it is also conceivable that $3\beta/2A$ approaches finite or even vanishing values at the endpoints of some intervals. Such behavior that depends on the side of approach can also be consistent with essential singularities.

We now turn to a $SU(N_c)$ gauge theory, where results are expressed in terms of the Casimirs, $C_G=N_c$, $C_R=(N_c^2-1)/2N_c$, $T_R=1/2$. $N$ still denotes the number of flavors. The term in the $\beta$-function of zeroth order in the $1/N$ expansion is as before, $\beta(\alpha)=2A/3+...$, if we choose a new definition for $A\equiv N T_R\alpha/\pi$. Similarly the $1/N$ expansion is
\begin{equation}
\frac{3}{2}\frac{\beta(\alpha)}{A}=1+\sum_{i=1}^{\infty}\frac{H_i(A)}{N^i}.
\label{e8}\end{equation}
From the results in \cite{a10} we can deduce that\footnote{This result is presented somewhat more explicitly than in \cite{a10}, but its series expansion is in agreement.}
\begin{eqnarray}
H_1(A)&=&-\frac{11}{4}\frac{C_G}{T_R}+\int_0^{A/3} I_1(x)I_2(x)dx,\label{e7}\\
I_2(x)&=&\frac{C_{{R}}}{T_R}+\frac{\left( 20-43\,x+32\,{x}^{2}-14\,{x}^
{3}+4\,{x}^{4} \right) }{ 4\left( 2\,x-1 \right)  \left( 2\,x-3 \right) \left( 1-x^2
 \right) }\frac{C_{{G}}}{T_R}.
\end{eqnarray}
$I_1(x)$ is defined in (\ref{e9}) and so up to the definition of $A$ and the $C_R/T_R$ factor, the $C_R$ term is just the QED result for $F_1(A)$. By inspection of $I_1(x)$ and $I_2(x)$ one can see that the $C_G$ term brings in a new pole in the integrand at $x=1$ ($A=3$). This pole is of the same sign as the pole at $A=15/2$. We show a plot of $H_1(A)$ in Fig.~(4) for $N_c=3$. Compared to the $F_1(A)$ function of QED, $H_1(A)$ is negative and its first singularity is occurring at a smaller value of $A$ (and $\alpha$).
\begin{figure}[t]
\centering\includegraphics[scale=0.5]{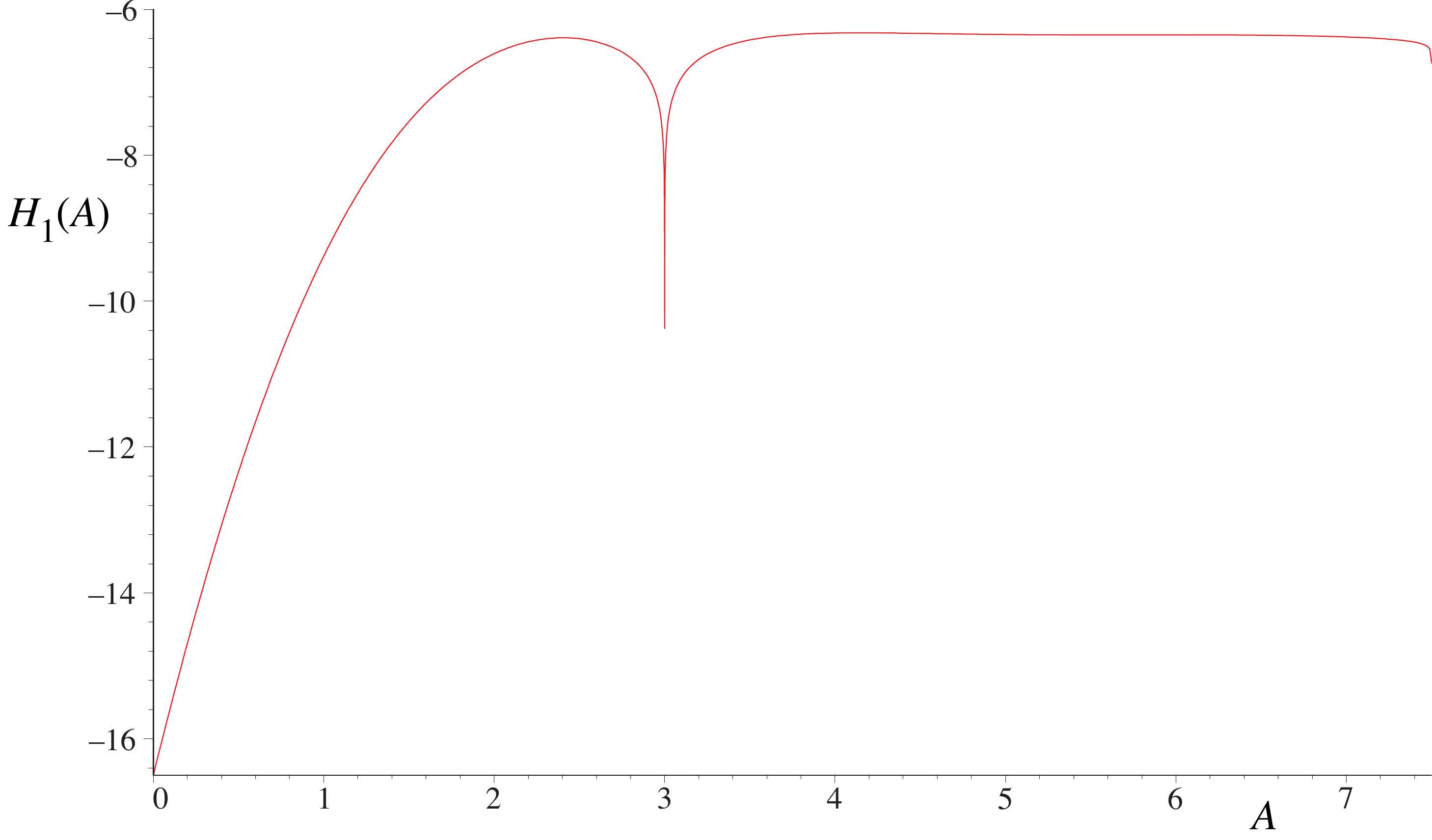}
\caption{$H_1(A)$ as defined in (\ref{e7}) for $N_c=3$.}\end{figure}

As in the QED case, a key question is how large $N$ has to be for the $1/N$ expansion to be under control, for $A$ at least as large as $3$. It is apparent that higher $N$ will be needed, since factors of $N$ must now compete with factors of $N_c$. If we rescale $A=3\tilde{A}$ and $N=32\tilde{N}$ then the expansion (\ref{e8}) for $N_c=3$ numerically reads
\begin{eqnarray}
1&+&\frac{1}{\tilde{N}}(-0.5156+0.8906\tilde{A}-.6348\tilde{A}^2-.2372\tilde{A}^3+.4372\tilde{A}^4-
0.0994\tilde{A}^5-.0912\tilde{A}^6+...)\nonumber\\
&+&\frac{1}{\tilde{N}^2}(-0.2241\tilde{A}+0.9216\tilde{A}^2-2.003\tilde{A}^3+...)\\
&+&\frac{1}{\tilde{N}^3}(-.1471\tilde{A}^2+1.073\tilde{A}^3+...)+\frac{1}{\tilde{N}^4}(-.1412\tilde{A}^3+...)\nonumber
\end{eqnarray}
The leading terms of the expansions of $H_2(A)$, $H_3(A)$, $H_4(A)$ have been obtained from the 4-loop results in \cite{a11}. Due to the pure glue contributions, each order begins with a lower power of $A$ as compared to QED. The main observation here is that large values of $N$ are needed, $N\gtrsim32$ or more, for the $1/N$ expansion to be under control.
\begin{figure}[t]
\centering\includegraphics[scale=0.5]{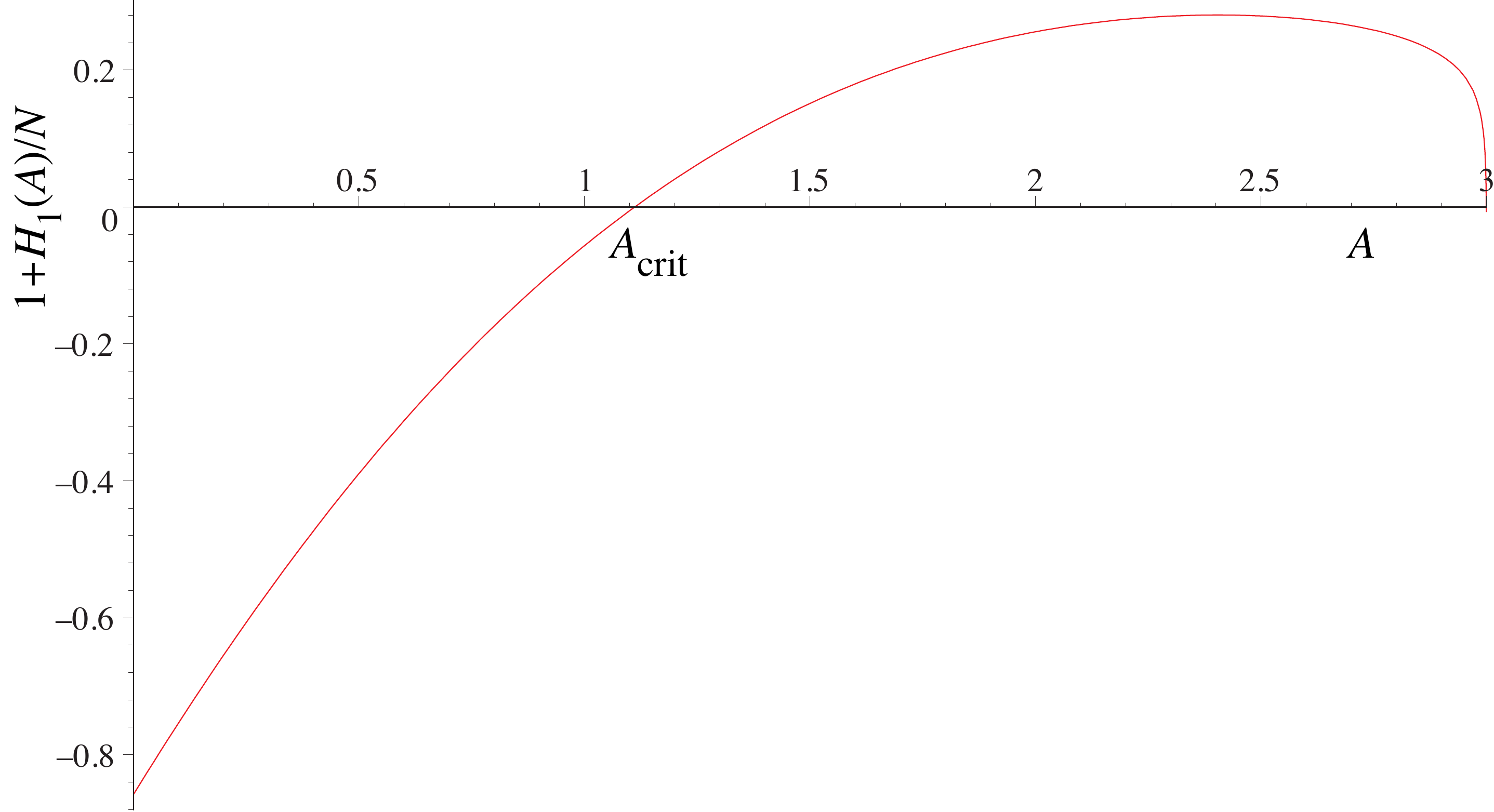}
\caption{$3\beta(\alpha)/2A$ at first order in $1/N$ where a zero occurs at the critical coupling for chiral symmetry breaking ($N=8.88$ and $N_c=3$).}\end{figure}

For such large values of $N$, asymptotic freedom has been lost, and the implications of the singularities in $H_1(A)$ and the higher $H_i(A)$'s will resemble that of QED. The large $N$ required also indicates that the large $N$ expansion lacks quantitative control for the study of other interesting phenomena that occur for values of $N\ll32$.\footnote{Of course if the study involved values of $A$ significantly less than $A=3$ then such large values of $N$ would not be needed.} One such question is the lowest value of $N$ at which the Banks-Zaks fixed point survives before chiral symmetry breaking occurs.

For completeness we give the answer to this question while only keeping the $1+H_1(A)/N$ terms in (\ref{e8}). We can find the $N$ such that the zero of $1+H_1(A)/N$ occurs at the usual estimate for the critical coupling for chiral symmetry breaking, or $A_{\rm crit}=N T_R/3C_R$. For $N_c=3$ this gives $N=8.88$ as the lower bound on the conformal window. This may be compared with the conventional result of $N=11.9$ which is derived from the 1- plus 2-loop terms of $\beta(\alpha)$. We plot $1+H_1(A)/N$ for $N=8.88$ in Fig.~(5).

At order $1/N^2$, the singularity structure of the QCD $\beta$-function is significantly more complicated. And unlike QED, the singularity structures of the $\gamma_m$ and $\beta$-functions do not match. For example $H'_1(A)$ has the additional singularity at $A=3$ when compared to the QCD $\gamma_m$-function at order $1/N$, which is essentially the QED result $G_1(A)$. As mentioned before, this can be understood diagrammatically in terms of additional contributions to the $\beta$-function due to the gluon self-coupling. Nevertheless there will be contributions to singularities in $H'_2(A)$ that can be associated with the singularities that appear at order $1/N^2$ in the QCD $\gamma_m$-function. Since the latter is known \cite{a13} we can deduce that the $H_2(A)$ singularity structure will contain the following contributions: a simple UV pole at $A=3/2$, a positive log singularity at $A=3$,\footnote{This would be subdominant to the negative log singularity appearing in $H_1(A)$.} a negative log singularity at $A=9/2$, and IR poles of third order at $A=15/2+n$. Some of these singularities could change sign or be replaced by even stronger singularities in the full result for $H_2(A)$. The UV pole is interesting since it would imply an infrared fixed point just above $A=3/2$.

In summary we have considered the implications of singularities appearing in the large $N$ flavor expansion of the $\beta$-functions of QED and QCD. The logarithmic singularities that appear at first order in $1/N$ are the first signs of simple and higher order poles that will appear at higher orders in $1/N$. In the case of QED these singularities are expected to be related to those of the $\gamma_m$-function, for which the singularity structure is surprisingly simple at order $1/N^2$. This gives information about the signs and locations of a set of poles in the $\beta$-function at this order. This in turn implies an infinite set of nontrivial infrared fixed points. The poles are likely to turn into essential singularities when all orders of the $1/N$ expansion are considered, but the existence of nontrivial fixed points will persist. For QCD we have indicated an even richer singularity structure for which we have less information.

\section*{Acknowledgments}
I am grateful to J.~Gracey for bringing to my attention his results in \cite{a10}, which extends the $1/N$ expansion to the QCD $\beta$-function. This work was supported in part by the Natural Science and Engineering Research Council of Canada.

\end{document}